\documentstyle[preprint,eqsecnum,aps,epsf]{revtex}

\begin{document}
\preprint{\vbox{
\hbox{INPP-UVA-97-07} 
\hbox{December 1997} 
\hbox{hep-ph/9801332}
}}
\draft
\def\be{\begin{eqnarray}}
\def\en{\end{eqnarray}}
\def\non{\nonumber}
\def\la{\langle}
\def\ra{\rangle}
\def\up{\uparrow}
\def\dw{\downarrow}
\def\ep{\varepsilon}
\def\ms{\overline{\rm MS}}
\def\ums{{\mu}_{_{\overline{\rm MS}}}}
\def\u{\mu_{\rm fact}}
\def\pr{{\sl Phys. Rev.}~}
\def\ijmp{{\sl Int. J. Mod. Phys.}~}
\def\mpl{{\sl Mod. Phys. Lett.}~}
\def\prp{{\sl Phys. Rep.}~}
\def\prl{{\sl Phys. Rev. Lett.}~}
\def\pl{{\sl Phys. Lett.}~}
\def\np{{\sl Nucl. Phys.}~}
\def\ppnp{{\sl Prog. Part. Nucl. Phys.}~}
\def\zp{{\sl Z. Phys.}~}

\title{\bf Quark Orbital Motion in the Nucleon\\}

\author{Xiaotong Song\\}

\address{Institute of Nuclear and Particle Physics\\
Department of Physics, University of Virginia\\
Charlottesville, VA 22901, USA\\}
\date{Nov.15, 1997. Revised Dec. 20, 1997 and Jan. 21, 1997 }
\maketitle
\vskip 12pt
\begin{abstract}
An unified scheme for describing both spin and orbital 
motion in symmetry-breaking chiral quark model is suggested. 
The analytic results of the spin and orbital angular momenta 
carried by different quark flavors in the nucleon are given.
The quark spin reduction due to spin-flip in the chiral 
splitting processes is compensated by the increase of the 
orbital angular momentum carried by the quarks and antiquarks. 
The sum of both spin and orbital angular momenta in the nucleon 
is 1/2, if the gluons and other degrees of freedom are neglected. 
The same conclusion holds for other octet and decuplet baryons. 
Possible modification and application are briefly discussed.
\end{abstract}

\bigskip
\bigskip
\bigskip


\newpage


In the past decade, considerable experimental and theoretical 
progress has been made in determining the quark spin contribution 
in the nucleon \cite{review}. The polarized deep-inelastic scattering
(DIS) data \cite{emc,smc-slac-hermes,smc97} indicate that the quark spin
only contributes about one third of the nucleon spin. A natural and 
interesting question is where is the {\it missing} spin ? The nucleon 
spin can be decomposed into three gauge-invariant pieces \cite{ji}  
$$
{1\over 2}~=~{1\over 2}\Delta\Sigma~+~<L_z>_{q+\bar q}~+~<J_z>_{G}
\eqno (1)
$$
without loss of generality, the proton is chosen to be {\it longitudinal 
polarized} in $z$ direction and has helicity of +${1\over 2}$. 
${1\over 2}\Delta\Sigma={1\over 2}\sum\limits[\Delta q+\Delta\bar q]$ 
is the total spin from quarks and antiquarks. $\Delta q\equiv 
q_{\up}-q_{\dw}$ and $\Delta{\bar q}\equiv{\bar q}_{\up}-{\bar q}_{\dw}$,
where $q_{\up,\dw}$ (${\bar q}_{\up,\dw}$) are quark (antiquark) 
{\it numbers} of spin parallel and antiparallel to the nucleon spin,  
or more precisely, quark numbers of {\it positive} and {\it negative}
helicities. $<L_z>_{q+\bar q}$ denotes the total orbital angular 
momentum carried by {\it quarks and antiquarks}, and $<J_z>_{G}$ is the 
gluon angular momentum. The smallness of ${1\over 2}\Delta\Sigma$ implies 
that the missing part should be contributed by either the orbital 
motion or gluon angular momentum. Further separation of $<J_z>_G$ 
into the spin and orbital pieces $\Delta G$ and $<L_z>_G$ is gauge 
dependent. Recently, it has been suggested that $<J_z>_{q+\bar q}=
{1\over 2}\Delta\Sigma+<L_z>_{q+\bar q}$ can be measured in the deep
virtual compton scattering process \cite{ji-jms}.

In the naive quark model \cite{gellmann64}, all three quarks in the 
nucleon are assumed to be in S-states, so $<L_z>_q=0$ and the nucleon 
spin is entirely attributed to the quark spin. On the other hand, in 
the naive parton model \cite{feynman69}, all quarks, antiquarks and 
gluons are moving in the same direction, i.e. parallel to the proton 
momentum, there is no transverse momentum for the partons and thus 
$<L_z>_{q+\bar q}=0$ and $<L_z>_G=0$. This picture cannot be $Q^2$ 
independent due to QCD evolution. In leading-log approximation, 
$\Delta\Sigma$ is $Q^2$ independent while the gluon helicity $\Delta G$
increases with $Q^2$. This increase should be compensated by the 
decrease of the orbital angular momentum carried by partons (see for 
instance earlier paper \cite{bms79} and later analysis 
\cite{rat87-sd89-qrrs90}). Similar situation occurs in the spin  
reduction case in the chiral quark model as we will show below.
Recently, the leading-log QCD evolution of $<L_z>_{q+\bar q}$ and 
$<L_z>_G$ has been derived in \cite{jth96}. The perturbative QCD 
can predict $Q^2$ dependence of the spin and orbital angular momenta 
but not their values at the renormalization scale $\mu^2$, due to their 
nonperturbative origin. The chiral quark model may provides some
information on these quantities.

Phenomenologically, long before the EMC experimental data published 
\cite{emc}, using the Bjorken sum rule and low energy hyperon 
$\beta$-decay data, \cite{sehgal} shown that nearly $40\%$ of 
the nucleon spin arises from the orbital motion of quarks and rest 
$60\%$ is attributed to the spin of quarks and antiquarks. Most 
recently \cite{cs97} shown that to fit the baryon magnetic moments 
and polarized DIS data, a large collective orbital angular momentum
$<L_z>$, which contributes almost $80\%$ of nucleon spin, is needed. 
Hence the question of how much of the nucleon spin is coming from the
quark orbital motion remains. This paper will discuss this question 
within the chiral quark model.  

The chiral quark model was first formulated by Manohar and Georgi  
in \cite{mg} and describes successfully the nucleon properties in the 
scale range between $\Lambda_{\rm QCD}$ ($\sim$ 0.2-0.3 GeV) and 
$\Lambda_{\chi{\rm SB}}$ ($\sim$ 1 GeV). The dominant interaction is 
coupling among the constituent (dressed) quarks and Goldstone 
bosons, while the gluon effect is expected to be small. This model 
was first employed by Eichten, Hinchliffe and Quigg in \cite{ehq92} 
to explain both the {\it sea flavor asymmetry} and the {\it smallness of} 
$\Delta\Sigma$ in the nucleon. The model has been improved by 
introducing U(1)-breaking \cite{cl1} and kaonic suppression \cite{smw}. 
A complete description with both SU(3) and U(1)-breakings was developed 
in \cite{song9605}, (similar version was given in \cite{cl2},
another version with $\lambda_8$-breaking was given in \cite{wsk}), 
and has been reformed into an one-parameter scheme in \cite{song9705}.

The effective Lagrangian describing interaction between quarks and the 
octet Goldstone bosons and singlet $\eta'$ is 
$${\it L}_I=g_8{\bar q}\pmatrix{
{1\over {\sqrt 2}}{\pi}^{\rm o}
+{{\sqrt{\epsilon_{\eta}}}\over {\sqrt 6}}{\eta}
+{{\zeta'}\over {\sqrt 3}}\eta'
& {\pi}^+ & {\sqrt\epsilon}K^+\cr 
{\pi}^-& -{1\over {\sqrt 2}}{\pi}^{\rm o}
+{{\sqrt{\epsilon_{\eta}}}\over {\sqrt 6}}{\eta}
+{{\zeta'}\over {\sqrt 3}}\eta'
& {\sqrt\epsilon}K^{\rm o}\cr
{\sqrt\epsilon}K^-& {\sqrt\epsilon}{\bar K}^{\rm o}
& -{{2\sqrt{\epsilon_{\eta}}}\over {\sqrt 6}}{\eta}
+{{\zeta'}\over {\sqrt 3}}\eta'
\cr }q, 
\eqno (2)$$
where breakings are explicitly included. $a\equiv|g_8|^2$ denotes 
the transition probability of chiral fluctuation or splitting 
$u(d)\to d(u)+\pi^{+(-)}$, and $\epsilon a$ denotes the probability 
of $u(d)\to s+K^{-(0)}$. Similar definitions are used for 
$\epsilon_\eta a$ and $\zeta'^2a$. 

The basic assumptions of the chiral quark model we used are: 
(i) the nucleon flavor, spin and orbital contents are determined 
by its valence quark structure and all possible chiral fluctuations 
$q\to q'+{\rm GB}$, (ii) the coupling between the quarks and Goldstone
bosons is rather weak, one can treat the fluctuation $q\to q'+{\rm GB}$ 
as a small perturbation ($a\sim 0.10-0.15$) and the contributions 
from the higher order fluctuations can be neglected ($a^2<<1$), 
and (iii) the {\it valence quark structure} is assumed to be 
${\rm SU}(3)_{flavor}\otimes {\rm SU}(2)_{spin}$. Possible modification
on the third assumption will be discussed later.

The important features of the chiral fluctuation are that: (i) Due 
to the pseudoscalar coupling between the quarks and GB's, a quark 
{\it flips} its spin and changes (or maintains) its flavor by emitting 
a charged (or neutral) Goldstone bosons. The light quark sea asymmetry 
$\bar u<\bar d$ is attributed to the existing {\it flavor asymmetry of 
the valence quark numbers}, two valence $u$-quarks and one valence 
$d$-quark, in the proton. (ii) The quark spin reduction is due to the 
spin-flip in the chiral splitting processes $q_{\up}\to q_{\dw}+GB$. 
(iii) Most importantly, since the quark helicity flips in 
the fluctuations with GB emission, hence the quark spin component 
changes one unit of angular momentum, $(s_z)_f-(s_z)_i=+1$ or $-1$, 
the angular momentum conservation requires the {\it same amount change} 
of the orbital angular momentum but with {\it opposite sign}, i.e.
$(L_z)_f-(L_z)_i=-1$ or $+1$. This {\it induced orbital motion} 
distributes among the quarks and antiquarks, and compensates the
spin reduction in the dilution, and restores the angular momentum
conservation. This is the starting point to calculate the orbital 
angular momenta carried by quarks and antiquarks in the chiral 
quark model. 

For a spin-up valence $u$-quark, the allowed fluctuations are
$$u_{\up}\to d_{\dw}+\pi^+,~~
u_{\up}\to s_{\dw}+K^+,~~
u_{\up}\to u_{\dw}+({\rm GB}_+)^0,~~
u_{\up}\to u_{\up},
\eqno (3)$$
the $({\rm GB}_{\pm})^0$ denotes
$\pm {\pi^0}/{\sqrt 2}+{\sqrt{\epsilon_{\eta}}}{\eta^0}/{\sqrt 6}+
{\zeta'}{\eta'^0}/{\sqrt 3}$. Similarly, one can write down the allowed
fluctuations for $u_{\dw}$, $d_{\up}$, and $d_{\dw}$. Considering the 
valence quark {\it numbers} in the {\it proton}
$$n_p^{(v)}(u_{\up})={5\over 3}~,~~~n_p^{(v)}(u_{\dw})={1\over 3}~,~~~
n_p^{(v)}(d_{\up})={1\over 3}~,~~~n_p^{(v)}(d_{\dw})={2\over 3}~.
\eqno (4)$$
the spin-up and spin-down quark (or antiquark) contents, up to first
order chiral fluctuation, can be written as
$$n_p(q'_{\up,\dw}, {\rm or}\ {\bar q'}_{\up,\dw}) 
=\sum\limits_{q=u,d}\sum\limits_{h=\up,\dw}
n_p^{(v)}(q_h)P_{q_h}(q'_{\up,\dw}, {\rm or}\ {\bar q'}_{\up,\dw})
\eqno (5)$$
where $P_{q_{\up,\dw}}(q'_{\up,\dw})$ and $P_{q_{\up,\dw}}({\bar
q}'_{\up,\dw})$ are the probabilities of finding a quark $q'_{\up,\dw}$
or an antiquark $\bar q'_{\up,\dw}$ from all chiral fluctuations of a
valence quark $q_{\up,\dw}$, and can be obtained from the effective 
Lagrangian (2). They are listed in Table I, where 
$f\equiv{1\over 2}+{{\epsilon_{\eta}}\over 6}+{{\zeta'^2}\over 3}$,
$A\equiv 1-\zeta'+{{1-{\sqrt\epsilon_{\eta}}}\over 2}$, and
$B\equiv \zeta'-{\sqrt\epsilon_{\eta}}$. Using (4), (5) and Table
I, we can obtain all quark (antiquark) flavor and spin contents in 
the proton \cite{song9705}. Especially, 
$$\Delta u^p={4\over 5}\Delta_3-a,~~\Delta d^p=-{1\over 5}\Delta_3-a,~~
\Delta s^p=-\epsilon a,
\eqno (6)$$
where $\Delta_3\equiv {5\over 3}[1-a(\epsilon+2f)]$.

The discussion of the orbital angular momentum contents is somewhat
different from above, because {\it only quark spin-flip fluctuations 
can induce change of the orbital angular momentum}.
For a spin-up valence $u$-quark, these fluctuations are {\it first 
three processes} in (3). The last process, $u_{\up}\to u_{\up}$ 
(means {\it no chiral fluctuation}) makes {\it no contribution} to 
the orbital motion, and will be disregarded. 
We assume that the orbital angular momentum produced from the 
splitting $q_{\up}\to q'_{\dw}+{\rm GB}$ is {\it equally shared} by 
all quarks and antiquarks, and introduce a {\it partition factor} $k$, 
which depends on the numbers of final state particles and interactions 
among them. If the Goldstone boson has a simple quark structure, i.e. 
each boson consists of a quark and an antiquark, one has {\it two 
quarks and one antiquark} (total number is {\it three}) after each 
splitting. Hence up to first order chiral fluctuation, one has 
$k=1/3$, if the interactions between the fluctuated quark and 
spectator quarks are neglected.

We define $<L_z>_{q'/q_{\up}}$ ($<L_z>_{{\bar q'}/q_{\up}}$) 
as the orbital angular momentum carried by the quark $q'$ 
(antiquark $\bar q'$), arises from a valence spin-up quark 
fluctuates into all allowed final states except for no emission 
case. Considering the quark spin component changes 
one unit of angular momentum in each splitting, 
we can obtain all $<L_z>_{q'/q_{\up}}$ and $<L_z>_{\bar q'/q_{\up}}$ 
for $q=u,d$ in the proton. Note
that $<L_z>_{q'/q_{\dw}}=-<L_z>_{q'/q_{\up}}$, and
similar relation holds for $\bar q'$. 
The total orbital angular momentum carried by a specific quark flavor, 
for instance $u$-quark in the proton, is
$$<L_z>_{u}^p=\sum\limits_{q=u,d}
[n_p^{(v)}(q_{\up})-n_p^{(v)}(q_{\dw})]<L_z>_{u/q_{\up}}
\eqno (7)$$
where $\sum$ summed over {\it valence quarks} $u$ and $d$ in
the proton. Similarly, one can obtain the $<L_z>_{d}^p$, 
$<L_z>_{s}^p$, and corresponding quantities for the antiquarks, 
they are listed in Table II. 

Defining $<L_z>_{q}^p$ ($<L_z>_{\bar q}^p$) as the total orbital 
angular momentum carried by all {\it quarks} ({\it antiquarks}), and 
we finally obtain
$$<L_z>_{q}^p=2ka(1+\epsilon+f),~~~<L_z>_{\bar q}^p=ka(1+\epsilon+f)
\eqno (8a)$$
$$<L_z>_{q+\bar q}^p\equiv<L_z>_{q}^p+<L_z>_{\bar q}^p=3ka(1+\epsilon+f)
\eqno (8b)$$
On the other hand, from (6), one has
$${1\over 2}\Delta\Sigma^p={1\over 2}-a(1+\epsilon+f)
\eqno (9)$$
The sum of (8b) and (9) gives
$$<J_z>_{q+\bar q}^p={1\over 2}-a(1-3k)(1+\epsilon+f)
\eqno (10)$$
Taking $k=1/3$, we obtain $<J_z>_{q+\bar q}^p=1/2$. This result shows 
that in the chiral fluctuations, the missing part of the quark spin 
{\it is transferred} into the orbital motion of quarks and antiquarks. 
The amount of quark spin reduction $a(1+\epsilon+f)$ in (9) is 
exactly canceled by the same amount increase of the quark orbital 
angular momentum in (8b), and the total angular momentum of 
nucleon is unchanged. This conclusion is {\it independent of the
probabilities of specific chiral fluctuations}. In addition, although
the orbital angular momentum carried by quarks (or antiquarks) 
$<L_z>_{q}^p$ (or $<L_z>_{\bar q}^p$) depends on the the chiral 
parameters, the ratio $<L_z>_{q}^p/<L_z>_{\bar q}^p=2$ is {\it 
independent of the probabilities of chiral fluctuations}. This is 
originated from the mechanism of the chiral splitting: there are 
{\it two} quarks and {\it one} antiquark after the splitting, and 
they equally share the orbital angular momentum produced in the 
splitting process. The total {\it loss} of quark spin $a(1+\epsilon+f)$ 
appeared in (9) is due to the fact that there are {three} splitting 
processes (for instance see (3)), which flip the quark spin, the
probabilities of these fluctuations are $a$, $\epsilon a$, and $fa$ 
respectively. The results for the proton hold for the neutron as well. 
The description given in this paper has been extended to other octet 
and decuplet baryons. The result shows that the loss of quark spin
due to spin-flip in the chiral splitting processes is compensated by 
the gain of the orbital angular momentum carried by the quarks and
antiquarks for {\it all baryons}. The sum of both spin and orbital 
angular momenta in the baryon is 1/2, if the gluons and other degrees 
of freedom are neglected. The detail results will be presented in a
forthcoming paper.

To see how much of the proton spin is contributed by the orbital 
motions, we first estimate (8a-b).
In full U(3) symmetry case, $1+\epsilon+f=3$, while in extreme 
SU(3)- and U(1)-breaking case ($\epsilon=\epsilon_{\eta}=\zeta'^2=0$),
$1+\epsilon+f=1.5$. The reality is presumably in between. The detail
analysis given in \cite{song9705} leads to $1+\epsilon+f\simeq 2.0$ 
and $a\simeq 0.15$, hence $<L_z>_{q}^p\simeq 0.20$ and 
$<L_z>_{\bar q}^p\simeq 0.10$. The orbital motions shared by different
quark flavors are listed in Table III, and compared with other models. 
Hence we have 
$<s_z>_{q+\bar q}^p/<J_z>_{q+\bar q}^p\simeq 2/5$, and 
$<L_z>_{q+\bar q}^p/<J_z>_{q+\bar q}^p\simeq 3/5$. i.e. nearly 
$60\%$ of the proton spin is coming from the orbital motion of
quarks and antiquarks, and $40\%$ is contributed by the quark spin. 
The ratio of the spin to orbital angular momenta is 
$<s_z>_{q+\bar q}^p/<L_z>_{q+\bar q}^p \simeq 2/3$.

We have assumed that there are {\it no gluons} and other degrees 
of freedom in the proton, hence $<J_z>_G$=0. This is presumably
a good approximation at very low $Q^2$. However, if $<J_z>_G$ is 
nonzero \cite{bs90} and not small, the results given above should 
be modified. Taking $<J_z>_G(1~{GeV^2})\simeq 0.25\pm 0.10$ given 
in \cite{bj97}, and {\it assuming} the ratios derived from the 
chiral quark model still hold, one has $<L_z>_{q+\bar q}(1~{GeV^2})
\simeq 0.15\pm 0.07$ and $<s_z>_{q+\bar q}(1~{GeV^2})\simeq 0.10
\pm 0.07$, which is consistent with DIS data 
\cite{smc-slac-hermes,smc97}, and lattice QCD result \cite{dll95}.

One of important applications of our description is to study the 
baryon magnetic moments, which should depend on both spin and orbital 
motions of quarks and antiquarks. Assuming $\mu_u=-2\mu_d=-3\mu_s$,
the ratio of the proton to neutron magnetic moments is ${\mu_p}/{\mu_n}
=-(3/2)[1-(5a/6)(1-2\epsilon/3)/(1-a(2\xi'-2\epsilon/3-5/2))]$, 
where $\xi'\equiv(1-k)(1+\epsilon+f)$. If the orbital motion 
is not included ($k=0$), one obtains ${{\mu_p}/{\mu_n}}\simeq -1.33$,
while for $k=1/3$, ${{\mu_p}/{\mu_n}}\simeq -1.38$ (data: $-1.48$). 
The agreement with data is improved. A detail discussion of the baryon 
magnetic moments will be presented in another paper.

To summary, we have developed a new and unified scheme for describing 
both spin and orbital motions of quarks and antiquarks in symmetry
breaking chiral quark model. The orbital motions carried by different 
quark flavors in the proton are calculated. Extension, modification and
application of this scheme will be presented elsewhere.

The author thanks Xiangdong Ji and H. J. Weber for reading 
the manuscript and useful comments. This work was supported in part by 
the U.S. DOE Grant No. DE-FG02-96ER-40950, the Institute of Nuclear and 
Particle Physics, University of Virginia, and the Commonwealth of Virginia.

{\it Note added. $-$~} After this preprint was posted, the author learned 
that the basic feature of orbital angular momentum in the nucleon has 
been suggested before by Ta-Pei Cheng and Ling-Fong Li \cite{cl3}. 
However, our consideration is more general in the chiral quark model 
with SU(3) and U(1) breakings, and thus has different conclusions on 
the octet and decuplet baryons and their magnetic moments.


\begin{table}[ht]
\begin{center}
\caption{The probabilities $P_{u_{\up}}(q'_{\up,\dw},\bar q'_{\up,\dw})$
and 
$P_{d_{\up}}(q'_{\up,\dw},{\bar q}'_{\up,\dw})$} 
\begin{tabular}{ccc} 
$q'$ &$P_{u_{\up}}(q'_{\up,\dw})$ & $P_{d_{\up}}(q'_{\up,\dw})$ \\ 
\hline 
$u_{\up}$ & $1-({{1+\epsilon}\over 2}+f)a+
{a\over {18}}(3-A)^2$ & ${a\over {18}}A^2$  \\
$u_{\dw}$ & $({{1+\epsilon}\over 2}+f)a+
{a\over {18}}(3-A)^2$ & $a+{a\over {18}}A^2$ \\
$d_{\up}$ & ${a\over {18}}A^2$ &$1-({{1+\epsilon}\over 2}+f)a+
 {a\over {18}}(3-A)^2$\\
$d_{\dw}$ & $a+{a\over {18}}A^2$ &
$({{1+\epsilon}\over 2}+f)a+{a\over {18}}(3-A)^2$ \\
$s_{\up}$ & ${a\over {18}}B^2$ &${a\over {18}}B^2$ \\
$s_{\dw}$ & $\epsilon a+{a\over {18}}B^2$ & $\epsilon a+{a\over {18}}B^2$
\\
\hline
${\bar u}_{\up,\dw}$ &${a\over {18}}(3-A)^2$ & ${a\over 2}+{a\over
{18}}A^2$\\
${\bar d}_{\up,\dw}$ &${a\over 2}+{a\over {18}}A^2$&
${a\over {18}}(3-A)^2$\\
${\bar s}_{\up,\dw}$ &${{\epsilon a}\over 2}+{a\over {18}}B^2$&
${{\epsilon a}\over 2}+{a\over {18}}B^2$\\
\end{tabular}
\end{center}
\end{table}

\begin{table}[ht]
\begin{center}
\caption{The orbital angular momentum carried by the quark $u$, $d$,
and $s$, and antiquark $\bar u$, $\bar d$, and $\bar s$ in the proton.}
\begin{tabular}{cc} 
$<L_z>_{u}^p={{ka}\over 3}[4(1+\epsilon+f)-1+{{4(3-A)^2}\over 9}
-{{A^2}\over 9}]$ & 
$<L_z>_{\bar u}^p={{ka}\over 3}[-1+{{4(3-A)^2}\over 9}-{{A^2}\over 9}]$ \\
\hline
$<L_z>_{d}^p={{ka}\over 3}[4-(1+\epsilon+f)+{{4A^2}\over 9}
-{{(3-A)^2}\over 9}]$ & 
$<L_z>_{\bar d}^p={{ka}\over 3}[4+{{4A^2}\over 9}-{{(3-A)^2}\over 9}]$ \\
\hline
$<L_z>_{s}^p={{ka}\over 3}[3\epsilon+{{B^2}\over 3}]$&
$<L_z>_{\bar s}^p={{ka}\over 3}[3\epsilon+{{B^2}\over 3}]$\\
\end{tabular}
\end{center}
\end{table}

\begin{table}[ht]
\begin{center}
\caption{Quark spin and orbital angular momenta in the chiral quark model 
and other models.}
\begin{tabular}{cccccc} 
Quantity & Data \cite{smc97}   & This paper &  Sehgal \cite{sehgal} 
&CS \cite{cs97}& NQM\\ 
\hline 
$<L_z>_u^p$       & $-$ & 0.136   &0.237   & $-$    & 0  \\ 
$<L_z>_{\bar u}^p$& $-$ & $-$0.002&0       & $-$    & 0  \\ 
$<L_z>_d^p$       & $-$ & 0.044   &$-$0.026& $-$    & 0  \\ 
$<L_z>_{\bar d}^p$& $-$ & 0.079   &0       & $-$    & 0  \\ 
$<L_z>_s^p$       & $-$ & 0.026   &0       & $-$    & 0  \\ 
$<L_z>_{\bar s}^p$& $-$ & 0.026   &0       & $-$    & 0  \\ 
\hline
$<L_z>_{q+\bar q}^p$ & $-$ & 0.31 &0.21  & 0.39    & 0  \\ 
\hline
$\Delta u^p$ & $0.85\pm 0.04$ & 0.85   & 0.91 & 0.78    & 2/3  \\ 
$\Delta d^p$ & $-0.41\pm 0.04$ & $-0.40$   & $-0.34$ & $-0.48$ & $-1/6$\\ 
$\Delta s^p$ & $-0.07\pm 0.04$ & $-0.07$   & 0 & $-0.14$ & 0  \\ 
\hline
${1\over 2}\Delta\Sigma^p$ & $0.19\pm 0.06$ & 0.19 & 0.29 & 0.08 & 1/2\\ 
\end{tabular}
\end{center}
\end{table}

\end{document}